# Anisotropic temperature dependence of normal state resistivity in underdoped region of a layered electron-doped superconducto $Nd_{2-x}Ce_xCuO_4$

A.S. Klepikova[1], T.B. Charikova[1,2], N.G. Shelushinina[1], M.R. Popov[1], A. A. Ivanov[3]

[1] M.N.Mikheev Institute of Metal Physics Ural Branch of RAS, Ekaterinburg, Russia
[2] Ural Federal University, Ekaterinburg, Russia
[3] National Research Nuclear University MEPhI, Moscow, Russia
E-mail: Popov_mr@imp.uran.ru

Abstract

The aim of this work is to investigate the temperature dependencies both in $CuO_2$ – plane and out-of plane resistivities in electron-doped $Nd_{2-x}Ce_xCuO_4$ for x from 0.135 up to 015 in order to analyze the anisotropy of the electrical transport in the process of the evolution from antiferromagnetic (AF) order in the underdoped region to superconducting (SC) order in optimally doped region.

1. Introduction

The problem of the resistivity anisotropy in the normal state of copper oxide systems has long attracted the attention of researchers. A strong anisotropy of the conducting properties ($\rho_c/\rho_{ab} \gg 1$) when a nonmetallic conductivity along the $c$ axis is combined with a metallic conductivity in the $ab$ plane was repeatedly observed in underdoped and optimally doped *hole-type* HTSCs [1,2]. This is evidence of the quasi-two-dimensionality of oxide systems that consist of highly mobile $CuO_2$ layers separated by buffer layers [3,4]. The nonmetallic character of $\rho_c$ in most superconducting high-$T$c compounds suggests an unconventional conduction mechanism between $CuO_2$ planes.

One of the most fundamental concepts in solid state physics is that in most metallic crystals the electronic conduction occurs through the coherent motion of electrons in band states associated with well-defined wave vectors [5]. There is currently a great deal of interest in whether this concept is valid for interlayer transport in high-$Tc$ superconductors [6]. Incoherent transport means that the motion from layer to layer is diffusive and band states and a Fermi velocity perpendicular to the layers cannot be defined. The Fermi surface is then not three-dimensional and Boltzmann transport theory cannot describe the interlayer transport [7].

The cerium-doped cuprate of $Nd_{2-x}Ce_xCuO_{4+\delta}$ has a layered quasi-two-dimensional perovskite-like crystal structure [4]. As compared to other cuprate superconductors, $Nd_{2-x}Ce_xCuO_{4+\delta}$ has many unique properties that make it an attractive subject for investigations. This is a superconductor with an *electron-type* conductivity whose structure contains a single $CuO_2$ plane per unit cell. In optimally annealed crystals, there are no apical oxygen atoms between neighboring conducting $CuO_2$ planes. Therefore, $Nd_{2-x}Ce_xCuO_4$ crystals have clearly pronounced two-dimensional properties.

In bulk $Nd_{2-x}Ce_xCuO_{4+\delta}$ single crystals, a very strong anisotropy of the resistivity is observed, $\rho_c/\rho_{ab} \sim 10^4$ [8-10], however, the nonmetallic temperature dependence of $\rho_c(T)$ is quite rare. This, apparently, is due to the special sensitivity of the transport properties of the Nd - system to the content of non-stoichiometric oxygen ($\delta$) and difficulties in achieving an optimal annealing regime ($\delta \to 0$) for bulk samples. On the other hand, single-crystal $Nd_{2-x}Ce_xCuO_4/SrTiO_3$ films (up to 500 nm thick) are well suited for different annealing procedures.

High-quality single-crystal $Nd_{2-x}Ce_xCuO_{4+\delta}$ / $SrTiO_3$ films with the *c* axis perpendicular (orientation (001) [11]) and parallel (orientation (1 ī 0) [12]) to the substrate plane were obtained and investigated by us earlier. In [13] a comparative analysis of $\rho_{ab}(T)$ and $\rho_c(T)$ temperature dependences for single-crystal films $Nd_{2-x}Ce_xCuO_{4+\delta}$ / $SrTiO_3$ with $x = 0.12$ (nonsuperconducting underdoped films), $x = 0.15$ (optimally doped films with maximal $T_c$) and $x = 0.17, 0.20$ (overdoped films) at orientations (001) and (1 ī 0) were done. An emphasis has been placed on the overdoped region, the "damping" region of superconductivity(gradual decrease in $T_c$, down to $T_c \to 0$ for $x \approx 0.22$) in comparison with optimal doping one.

In the Nd-system (the *n-type* HTSC) we have found a transition from a quasi-2D ($d\rho_c/dT < 0, d\rho_{ab}/dT > 0$) to 3D anisotropic system with metallic conductivity both in the *ab*-plane and in the *c*-axis direction ($d\rho_c/dT > 0, d\rho_{ab}/dT > 0$) with increasing of doping, $x$, as it was found previously in the La- system (the *p-type* HTSC) [2]. From the works [2, 13] the conclusion about the correlation of quasi two-dimensionality of the system with the implementation of superconductivity in copper oxide compounds can be done.

Thus the region of the appearance of superconductivity with increasing cerium doping (at $x \approx 0.13 \div 0.14$) remained unexplored. Presently, the advances in technology have allowed us to grow the high-quality $Nd_{2-x}Ce_xCuO_{4+\delta}$ / $SrTiO_3$ single-crystal films with $x = 0.135$ and $x = 0.145$ in which the *c* - axis was both normal ((100) films) and parallel ((1 ī 0) films) to the substrate plane to study the carrier charge transfer processes in the region of the antiferromagnetic (AFM) – superconducting (SC) quantum phase transition.

Study of the properties just of these samples under an optimal annealing in conjunction with the optimally doped ($x = 0.15$) samples is the subject of this work. We are the first to detect experimentally the $\rho_c(T)$ dependences with nonmetallic behavior for samples with $x = 0.135$ and $x = 0.145$) near the threshold values of $x$ for AFM – SC phase transition. A comparison of the results obtained for the two types of films allowed us to demonstrate the quasi-two-dimensional character of carrier transport in them.

2. Materials and method

The superconductor with an electronic conductivity type $Nd_{2-x}Ce_xCuO_{4+\delta}$ has a body-centered crystal lattice and corresponds to a tetragonal T'-phase. Lattice parameters: $a = b = 0.39$ nm, $c = 1.208$ nm. As a result of doping and annealing ($\delta \to 0$), the crystal structure is a set of $CuO_2$ conducting planes separated by a distance of $d = c/2 = 0.6$ nm in the direction of the *c* axis [14]. The compound has pronounced two-dimensional properties - including quasi-two-dimensional character of the carrier transport.

We have synthesized $Nd_{2-x}Ce_xCuO_{4+\delta}$ / $SrTiO_3$ epitaxial films with $x = 0.135, 0.145$ and $0.15$ by pulsed laser deposition [15, 16] of two types :

1. Orientation of the film (001) - the c-axis of the $Nd_{2-x}Ce_xCuO_{4+\delta}$ lattice is perpendicular to the $SrTiO_3$ – substrate plane.

2. Orientation of the film (1$\bar{1}$0) - the c-axis of the $Nd_{2-x}Ce_xCuO_{4+\delta}$ lattice is directed along the long side of the $SrTiO_3$ substrate.

In the process of pulsed laser deposition, an excimer KrF laser was used with a wavelength of 248 nm, with an energy of 80 mJ/pulse. The energy density at the target surface was 1.5 J/cm². The pulse duration was 15 ns, the repetition rate of pulses was from 5 to 20 Hz. Further, the synthesized films were subjected to heat treatment (annealing) under various conditions to obtain samples with a maximum superconducting transition temperature. X-ray diffraction analysis (Co-K radiation) showed that all films were of high quality and were single crystal.

The optimum annealing conditions were as follows:

- for the composition $x = 0.15$ ($T_c^{onset} = 23.5\ K, T_c = 22\ K$) $- t = 60\ min., T = 780^0 C, p = 10^{-5}\ Torr$;

- for the composition $x = 0.145$ ($T_c^{onset} = 15.7\ K, T_c = 10.7\ K$) $- t = 60\ min., T = 600^0 C, p = 10^{-5}\ Torr$;

- for the composition $x = 0.135$ ($T_c^{onset} = 13.7\ K, T_c = 9.6\ K$) $- t = 60\ min., T = 600^0 C, p = 10^{-5}\ Torr$.

The thickness of the films was 140-520 nm.

The temperature dependences of the resistivity for both types of $Nd_{2-x}Ce_xCuO_{4+\delta}$ / $SrTiO_3$ films were carried out in the Quantum Design PPMS 9 and in the solenoid "Oxford Instruments" (Center for Nanotechnologies and Advanced Materials, IFM UrB RAS). The electric field was always applied parallel to the $SrTiO_3$ – substrate plane. Depending on the type of samples we were able to measure the temperature dependences of the resistivity in the conducting planes of $CuO_2$ and between planes (along the c axis).

3. Experimental results and discussion

The results for both in-plane, $\rho_{ab}$ (for films with (001) orientation), and out-of-plane, $\rho_c$ (films with (1$\bar{1}$0) orientation), resistivities as functions of the temperature in the samples $Nd_{2-x}Ce_xCuO_4$ / $SrTiO_3$ with $x = 0.135$, $0.145$ and $0.15$, optimally annealed in vacuum, are shown in Fig.1.

Let us discuss in more detail the temperature dependences of $\rho_{ab}$ and $\rho_c$ and their relation in terms of the anisotropic model for a quasi-two-dimensional system with good metallic conductivity in $CuO_2$ planes in combination with incoherent transfer between the planes.

3.1 Temperature dependence of resistivity in $CuO_2$ planes for films $Nd_{2-x}Ce_xCuO_4/SrTiO_3$ (001).

It is seen from Fig.1 that the normal state conductivity in the ab - plane is metallic with a dominant quadratic temperature dependence of $\rho(T)$ at $T > (25 - 70)K$ for different samples up to room temperature. A manifestation of weak 2D localization effects with $\rho(T) \sim \ln T$ takes place at T < 50K for $x = 0.145$ and at T < 70K for $x = 0.135$.

The resistance in the $CuO_2$ plane is described by the standard formula [5, 17]

$$\rho_{ab} = \frac{m}{ne^2\tau}, \qquad (1)$$

where *n* is the concentration and τ is the relaxation time of the carriers. Let us represent the total scattering probability in the form $\frac{\hbar}{\tau} = \frac{\hbar}{\tau_0} + \frac{\hbar}{\tau_{in}}$, where $\hbar/\tau_0$ describes the elastic scattering probability due to impurities and $\tau_{in}(T)$ is the inelastic scattering time responsible for the temperature dependence of the in-plane resistivity. Then we have

$$\rho_{ab}(T) = \rho_{ab}(0) + \Delta\rho_{ab}(T) \qquad (2)$$

with $\rho_{ab}(0) = \frac{m}{ne^2\tau_0}$ being the residual resistivity and $\Delta\rho_{ab}(T) = \frac{m}{ne^2\tau_{in}(T)}$.

In a paper of Kontani et al. [18] the authors give the explanation of a summary of the experimental relations on quadratic temperature dependence of the normal state in-plane resistivity $\rho_{ab}(T)$ for NdCeCuO in the underdoped region from the standpoint of the nearly antiferromagnetic (AF) Fermi liquid. The typical spin-fluctuation theories (see [19] and references therein) give $\rho \sim CT^2$ and thus may reproduce the experimental results.

On the other hand, Seng et al. [20] made a conclusion that quadratic temperature dependence of the zero-field resistivity $\rho_{ab}(T)$ in the normal state of $Nd_{1.85}Ce_{0.15}CuO_{4-\delta}$, observed by them, is generated by electron-e1ectron (e-e) scattering in a two-dimensional metal, i.e. $\tau_{in} \equiv \tau_{ee}$.

For electron-electron scattering in a three-dimensional (3D) metal the $T^2$ dependence of the zero-field resistivity should take place [17]. For a two-dimensional (2D) metal the $T^2$ law is modified by a logarithmic correction [21] and the dependence of $\Delta\rho_{ab}(T)$ takes the form:

$$\Delta\rho_{ab}(T) = K(T/T_{ee})^2 \ln(T_{ee}/T). \qquad (3)$$

Seng et al. [20] and also Tsuei et al. [22], in a first step, fitted a $T^2$ law to their data on the normal state resistivity of $Nd_{1.85}Ce_{0.15}CuO_{4-\delta}$ films. But next they found that their experimental results are better described by Eqs (2) and (3) with the residual resistivity, $\rho_{ab}(0)$, the factor $K$ and the effective temperature, $T_{ee}$, as fitting parameters.

The solid lines in Fig. 1 are the best fits of Eqs (2) and (3) to our experimental data from 25K for sample with $x = 0.15$ (c), from 50K for sample with $x = 0.145$ (b) and from 70K for sample with $x = 0.135$ (a) up to room temperature with the parameters $\rho_{ab}(0)$, $K$ and $T_{ee}$ given in the Table 1.

The conclusion is that temperature dependence of zero-field resistivity $\rho_{ab}(T)$ in the normal state is generated by electron-electron scattering and that the good fit to the logarithmically corrected $T^2$ law (3) expresses the quasi-2D nature of the conductivity in our specimens.

3.2. The temperature dependence of resistivity across the $CuO_2$ planes for films $Nd_{2-x}Ce_xCuO_4 / SrTiO_3$ (1$\bar{1}$0).

It is seen from Fig. 1 that the normal state out-of-plane resistivity across the blocking layers, $\rho_c$, is large with respect to the in-plane resistivity, $\rho_{ab}$, and a non-metallic temperature dependence ($d\rho_c/dT < 0$) is observed for all the investigated films up to 300 K.

Two regions can be distinguished in the $\rho_c(T)$ dependence (see Fig. 1a,b,c): the high-temperature region, $T > 100K$, where $T$- dependence can be empirically described as $\rho_c(T) \sim 1/T$, and the low-temperature region, where out-of–plane resistivity exibits the activation- type temperature dependence.

The dashed lines in Figs 1 are the best fits of function $\rho(T) = a + b/T$ to the experimental data from 100K up to 300K for each sample. At $T < 100K$ the dependences of $ln\rho_c$ on $1/T$ may be described by straight lines corresponding to a function

$$\rho_c(T) = \rho_c^* exp\left(-\frac{\varepsilon_A}{kT}\right). \qquad (4)$$

with the values of activation energy $\varepsilon_A$ given in Table 1.

First the systematic data for $\rho_c(T)$ in a number of well characterized single high-$T_c$ crystals were presented by Ito et al.[1]. They find that $\rho_c(T)$ is non-metallic ($d\rho_c/dT < 0$) in most superconducting compounds, suggesting an unconventional conduction mechanism between $CuO_2$ planes in the normal state of superconducting cooper oxides.

The non-metallic $\rho_c(T)$ dependences that Ito et al. observed do not fit activation- or hopping-type laws, but exhibit the power law $T$ dependence, $\rho_c \infty - T^{-\alpha}$, with α in range $0 < \alpha < 2$. The authors arise a question whether the measured $\rho_c$ contains a contribution from $\rho_{ab}$ due to imperfect alignment of layers in the crystal.

Band calculations can explain the large anisotropy of resistivity in high-$T_c$ systems but predict that the out-of-plane conduction is always metallic, in sharp contrast to the experimental facts. As the simplest one-dimensional Kronig-Penny model with its ideal periodicity (and thus coherence) can only lead to a metallic nature of the interlayer conductivity, a number of microscopic models for deviation from coherence in c-axis transport have been proposed [7, 12, 23-27])

The effect of incoherent interlayer transport on the resistance of a layered metal was theoretically considered by McKenzie and Moses [7] wherein the Fermi surface appearance relevant to coherent and incoherent interlayer transport in a quasi-two-dimensional system was presented.

If the transport between layers is coherent then one can define a three-dimensional Fermi surface which is a warped cylinder. For the incoherent interlayer transport a Fermi surface is defined only within the layers ("a stack of pancakes") and the interlayer conductivity is determined by the interlayer tunneling rate (see FIG. 1 in [7]).

In [7, 25-27] the non-metallic behavior of $\rho_c(T)$ in the layered oxides was attributed to the *incoherent* tunneling of charge carriers in the *c*-axis direction. Incoherent transport between $CuO_2$ layers occurs when the probability of carrier scattering in the plane ($\hbar/\tau$) is much larger than the interlayer hopping integral $t_c (\equiv \hbar/\tau_{esc})$ between the planes. Here $\tau$ is the carrier relaxation time in the plane, and $\tau_{esc}$ is the escape time from the given plane to the neighboring one.

If an electron experiences many collisions before moving to another plane, the subsequent tunneling processes between the planes are uncorrelated. The interlayer conductivity, $\sigma_c^{tun}$, is then proportional to the tunneling rate between just two adjacent layers and can be represented in the following form (see, for example, [7, 25] and references therein):

$$\sigma_c^{tun} = 2de^2 g \left(\frac{t_c^2 \tau}{\hbar^2}\right), \qquad (5)$$

where $g = m/\pi\hbar^2$ is the density of states for unit area at the Fermi energy of the two-dimensional conducting planes.

In the simplified model of square barriers of the height $\Delta$, the tunneling matrix element $t_c$ can be calculated as

$$t_c = \exp(-d/r_0), \tag{6}$$

where $r_0$ is a radius of carrier localization, which is less than the distance between adjacent $CuO_2$ planes, and $r_0^{-1} = \sqrt{2m\Delta/\hbar^2}$.

Thus the $c$-axis resistivity for the tunneling process is found to be

$$\rho_c^{tun}(T) \equiv \frac{1}{\sigma_c^{tun}(T)} = A\rho_{ab}(T), \tag{7}$$

with $A = const$. The second equality in the right side of Eq. (7) can be obtained by expressing $1/\tau$ in terms of $\rho_{ab}$ using Eq. (1) (see [25] for more details on the relation of $\rho_c(T)$ and $\rho_{ab}(T)$).

Giura et al. [27] proposed the model for the $\rho_c(T)$ based on a submission that the crystal structure along the $c$ axis induces a stack of energy barriers between the nearly two-dimensional sheets where the carriers are mostly confined ($CuO_2$ layers). They assumed that two complementary processes determine the $c$-axis transport: incoherent tunneling and thermal activation across the barriers.

For the first mechanism Giura et al. adopted the model introduced in [7, 25 – 27] (see Eqs (5) or (7)) which describes the transport across a barrier trough incoherent tunneling process. For the second term, they used the general expression for thermal activation across the barrier:

$$\sigma_c^{th}(T) = \beta \cdot exp(-\Delta/kT), \tag{8}$$

where $\beta$ is a prefactor which may be weakly dependent on temperature.

The overall c-axis conductivity is then obtained as the combination of both contributions:

$$\sigma_c(T) = \sigma_c^{tun}(T) + \sigma_c^{th}(T). \tag{9}$$

In Fig. 1 we have described the behavior of $c$-axis resistivity at $T = (100 - 300)K$ by the empirical dependence $\rho_c(T) \sim 1/T$ in accordance with the Ito et al. approach [1]. As far as it is known no theoretical models have predicted such a behavior of the out-of-plane conduction in high-$T_c$ copper oxides.

We have attempted to describe the high-temperature part of $\rho_c(T)$ dependencies in the framework of the Giura et al. [27] model with the help of Eq.(9). Using expressions (7) and (8), we have found:

$$\sigma_c(T) = \alpha(T)exp(-q\sqrt{\Delta}) + \beta \cdot exp(-\Delta/kT), \tag{10}$$

with $\alpha(T) = \alpha_0/\rho_{ab}(T)$, $\alpha_0 = const$ and $q = 2d\sqrt{2m/\hbar^2}$, where the formula (6) for $t_c$ is used.

Figure 2 shows the experimental dependencies of $\sigma_c(T) \equiv 1/\rho_c(T)$, as well as the best fitting of these curves by the expression (10) with the adjustable parameters $\alpha_0$, $\beta$ and $\Delta$ (see Table 1).

The possibility of describing the experimental $\sigma_c(T)$ dependence at $T > 100K$ by the activation law (8) was verified on the insets of Fig. 2. It is seen that this law is valid only in the range of $T = (100 - 150)K$ for all the three samples with $\Delta = 8.1 meV$ for $x = 0.135$, $\Delta = 7.2 meV$ for $x = 0.145$ and $\Delta = 6.4 meV$ for $x = 0.15$.

On the other hand, the description of $c$-axis conductivity, $\sigma_c(T)$, by means of Eq. (10) is possible over a wide temperature ranges of $T = (150 - 300)K$ with $\Delta = 28.6 meV$ for $x = 0.135$, a of $T = (100 - 300)K$ with $\Delta = 17.4 meV$ for $x = 0.145$, and of $T = (75 - 250)K$ with $\Delta = 12.6 meV$ for $x = 0.15$ (see the main parts of Fig. 2a, b, c).

It is seen from Table 2 that the values of $\Delta$, found from the fitting procedure, decrease with increasing of $x$. This pattern is understandable if we take into account that, in the spirit of the Giura et al. model [27], the barrier height $\Delta$ in Eq. (8) is counted off from the Fermi level, $E_F$, in each system. The model can explain most of the qualitative features of observed resistivity, by assuming that an increase of $x$ results in an increase of the carrier density in the ($ab$)-planes and thus of the Fermi energy.

The continuous decrease of $\Delta$ as a function of $x$ also explains in a natural way the crossover from a semiconducting behavior of the $Nd_{2-x}Ce_xCuO_4$ normal-state resistivity at low doping ($x = 0.12 \div 0.15$) through almost metallic at slightly overdoped system ($x = 0.17$) to strictly metallic at highly overdoped region ($x = 0.20$) (see our work [12] for details)

At low temperatures ($kT << \Delta$) another mechanism of incoherent transfer between the layers can occur. In the model of a natural superlattice (when $CuO_2$ layers are quantum wells and $Nd(Ce)O$ blocks serve as barriers of the effective height $\Delta$) [27, 28, 29], we can consider the disorder, that is undoubtedly inherent in this system (the chaotic impurity potential), as a cause of the temperature dependence of $t_c$ [30].

Then in Eq. (5) we have

$$t_c \to t_c(T) = \exp(-d/r_0)\exp(-\delta/kT), \tag{11}$$

where $\delta$ is the spread of electron energy in the wells due to the fluctuations of $\Delta$ values, the same as in the one-dimensional Anderson model. The first factor in (11) (overlap integral) determines the dependence of the transition probability between the layers on the barrier height, and the second one leads to a nonmetallic temperature dependence of the conductivity at low temperatures (analog of the conductivity within the impurity band of semiconductors [30]). As $\sigma_c^{tun} \sim t_c^2$, in Eq. (4) the activation energy $\varepsilon_A = 2\delta$.

With increasing temperature, the contribution to the conductivity (8), associated with the thermal activation of carriers through the barrier, begins to play an increasingly important role and at $\delta \ll kT < \Delta$ we return to the mechanism of Giura et al. (see Eq.(10)).

3.3 Anisotropy of resistivity for optimally annealed $Nd_{2-x}Ce_xCuO_4/SrTiO_3$ films with various cerium contents.

For the 2D diffusion coefficients along ($D_\parallel$) and across ($D_\perp$) the layers (see [26], and references therein) one has: $D_\parallel = l^2/2\tau$ and $D_\perp = (d^2/2)(t_c/\hbar)^2\tau$, where $l$ is the mean free path in the $ab$ - plane. For the anisotropy coefficient of resistivity we then find:

$$\frac{\rho_c}{\rho_{ab}} = \frac{D_\parallel}{D_\perp} = (l/d)^2(\hbar/t_c\tau)^2 \tag{12}$$

and $\rho_c/\rho_{ab} \gg 1$, since $l/d \gg 1$ and under incoherent tunneling conditions $t_c\tau \ll \hbar$.

Thus, the strong anisotropy of the resistivity in the $Nd_{2-x}Ce_xCuO_{4+\delta}$ layered system can be explained by the incoherent transport of charge carriers in the $c$ direction with good metallic conductivity in the $CuO_2$ planes.

The insets of .Figs 1a,b,c show temperature dependencies of the resistivity anisotropy in the conducting $CuO_2$ planes and in a direction perpendicular to these planes for each of the films studied. It is seen that the coefficient of anisotropy of the resistivity is great even at room temperature: $\rho_c/\rho_{ab} \sim 10-10^2$. This parameter increases significantly with decreasing of $T$, reaching values $\rho_c/\rho_{ab} \sim 10^3$ for compounds with $x = 0.145$ and $0.135$ and $\rho_c/\rho_{ab} \sim 10^2$ for an optimally doped compound with $x = 0.15$, due to a sharp increase of $\rho_c$ at low temperatures.

We emphasize (see Fig.1) that low-temperature anisotropy coefficient is maximal for $x = 0.145$ ($\rho_c/\rho_{ab} \approx 2\cdot 10^3$) and $x = 0.135$ ($\rho_c/\rho_{ab} \approx 10^3$) and much less for $x = 0.15$ ( $\rho_c/\rho_{ab} \approx 10^2$) in contrast to the situation in the overdoped region where the direct correlation between the values of $\rho_c/\rho_{ab}$ and $T_c$ takes place (see Fig 6 in [12]).

**4.Conclusions**

The temperature dependencies of both in-plane ($\rho_{ab}$) and out-of-plane ($\rho_c$) resistivities in the normal state of recently grown $Nd_{2-x}Ce_xCuO_4/SrTiO_3$ structures with $x = 0.135$ and $0.145$, in the region of superconductivity emergence at the antiferromagnetic -superconductor transition boundary, have been studied. The results are compared with the data for the optimally doped superconducting $Nd_{2-x}Ce_xCuO_4$ structure with $x = 0.15$.

The structures are single-crystal $Nd_{2-x}Ce_xCuO_4/SrTiO_3$ films with the $c$-axis both perpendicular to the plane of the film (for measuring of $\rho_{ab}$) and parallel to the plane (for measuring of $\rho_c$).

The results obtained are successfully interpreted within the concept of quasi-two-dimensionality of the systems studied with high metallic conductivity along ($ab$) - planes ($d\rho_{ab}/dT > 0$) and semiconducting behavior of the conductivity in $c$ - axis direction ($d\rho_{ab}/dT < 0$) due to incoherent tunneling and thermal activation across the barriers between the conducting $CuO_2$ layers.

**Acknowledgments**

This work was carried out in the framework of the state task by the theme "Electron" (No. AAAA-A18-118020190098-5) and in part supported by the Russian Foundation for Basic Research (project No. 18-02-00192) and by the project "Fundamental Research" of Ural Branch of the Russian Academy of Sciences, No. 18-10-2-6.

Figures:

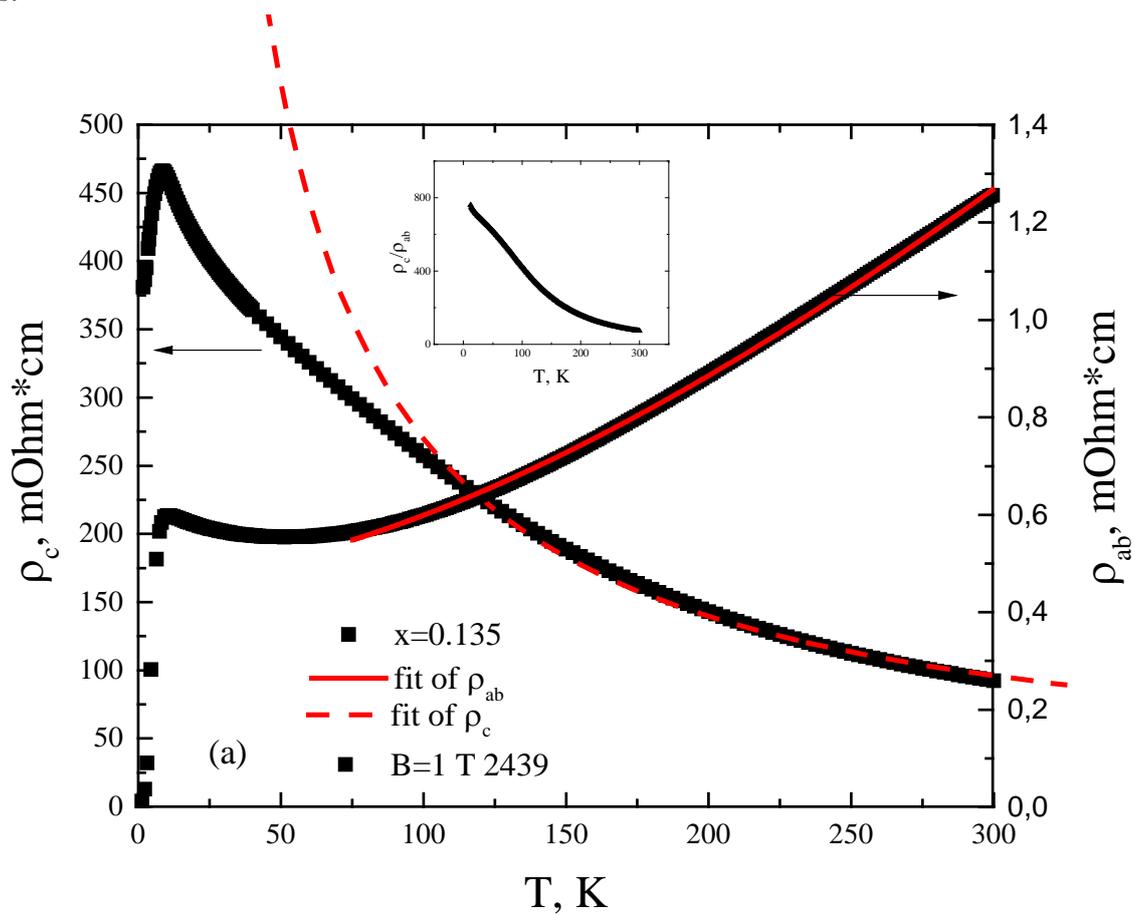

Fig1a

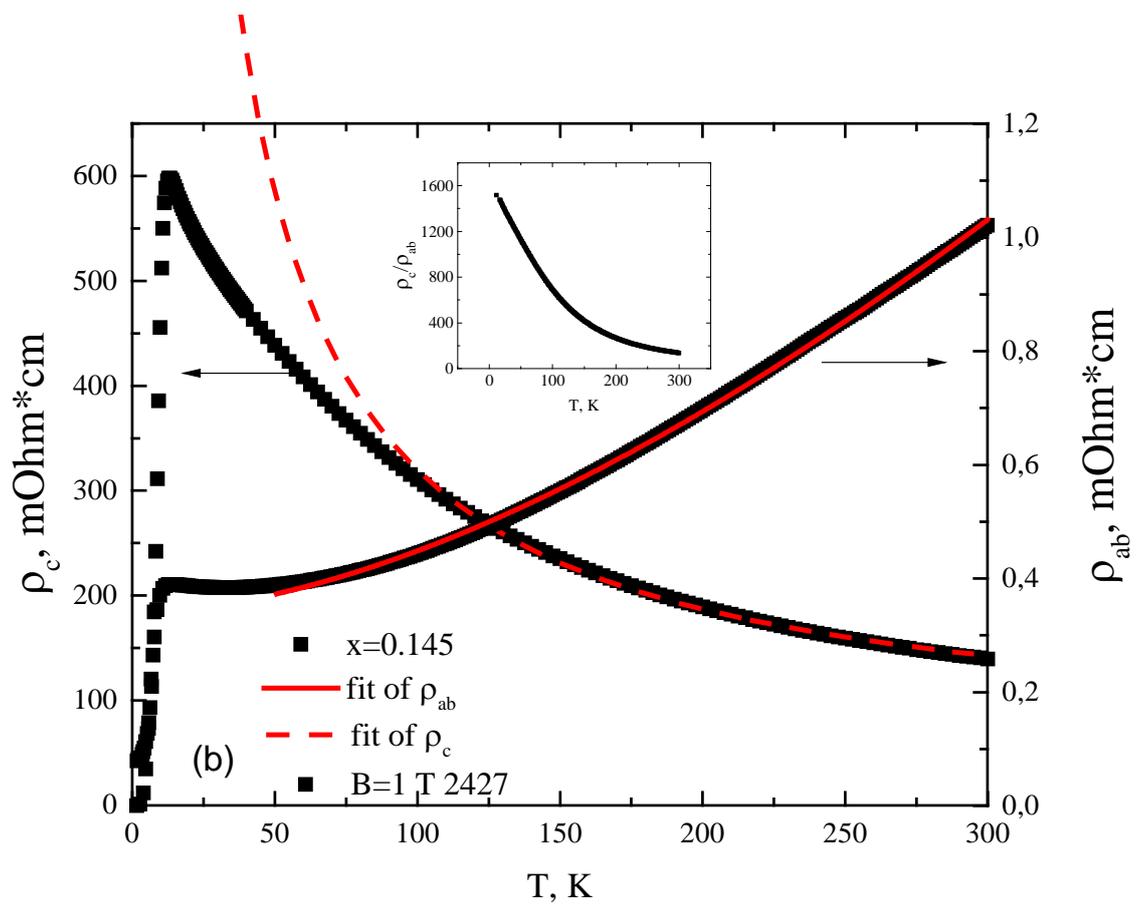

Fig1b

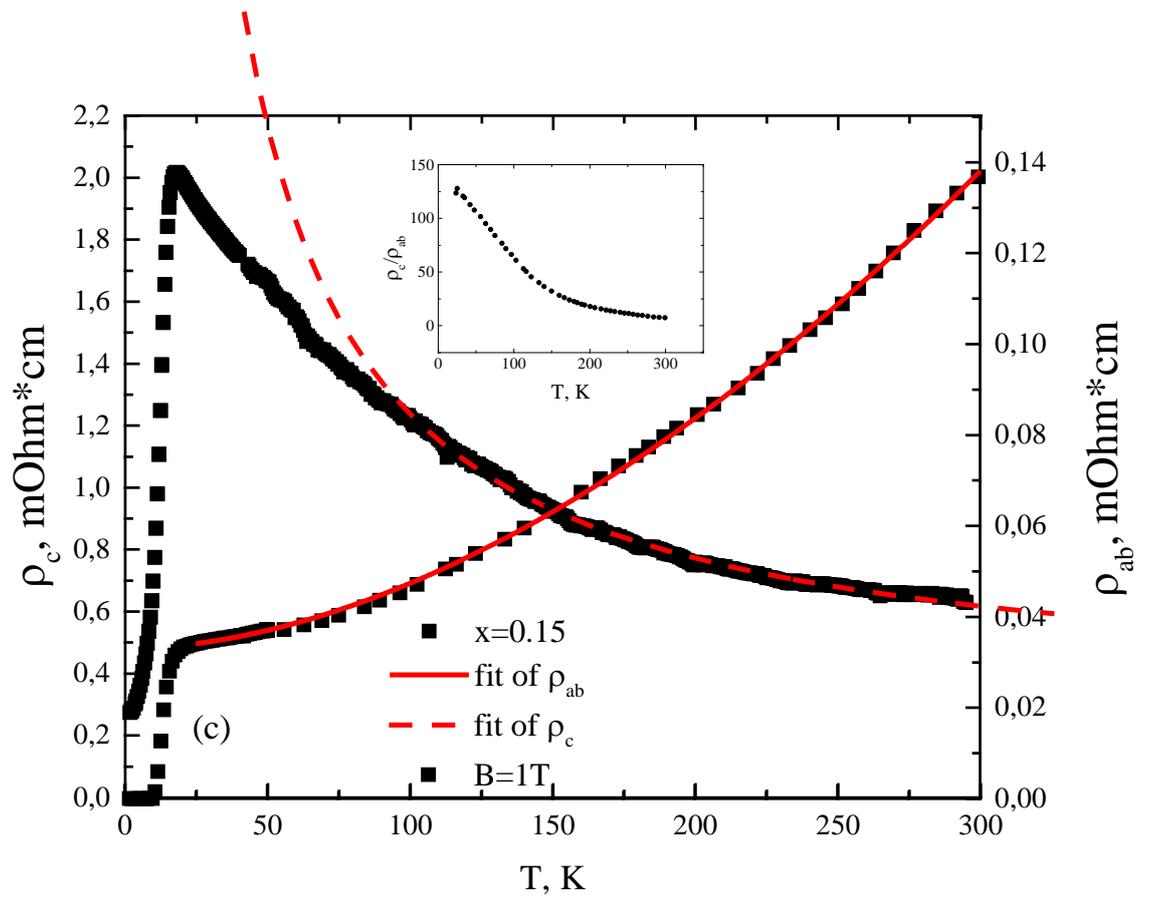

Fig1c

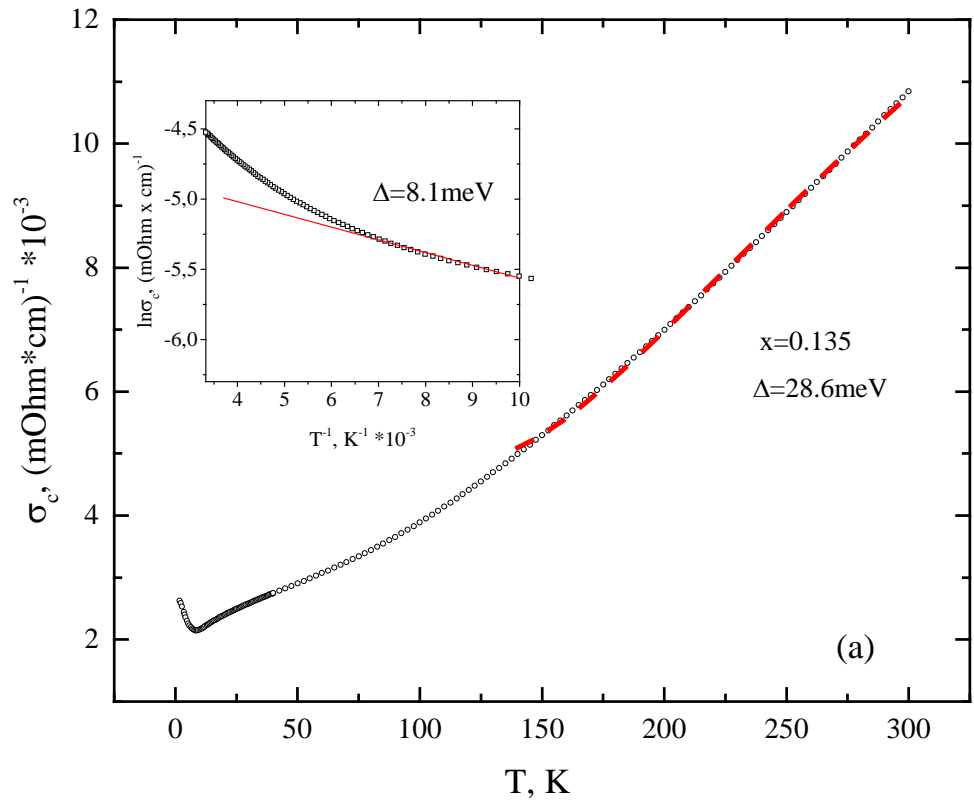

Fig2a

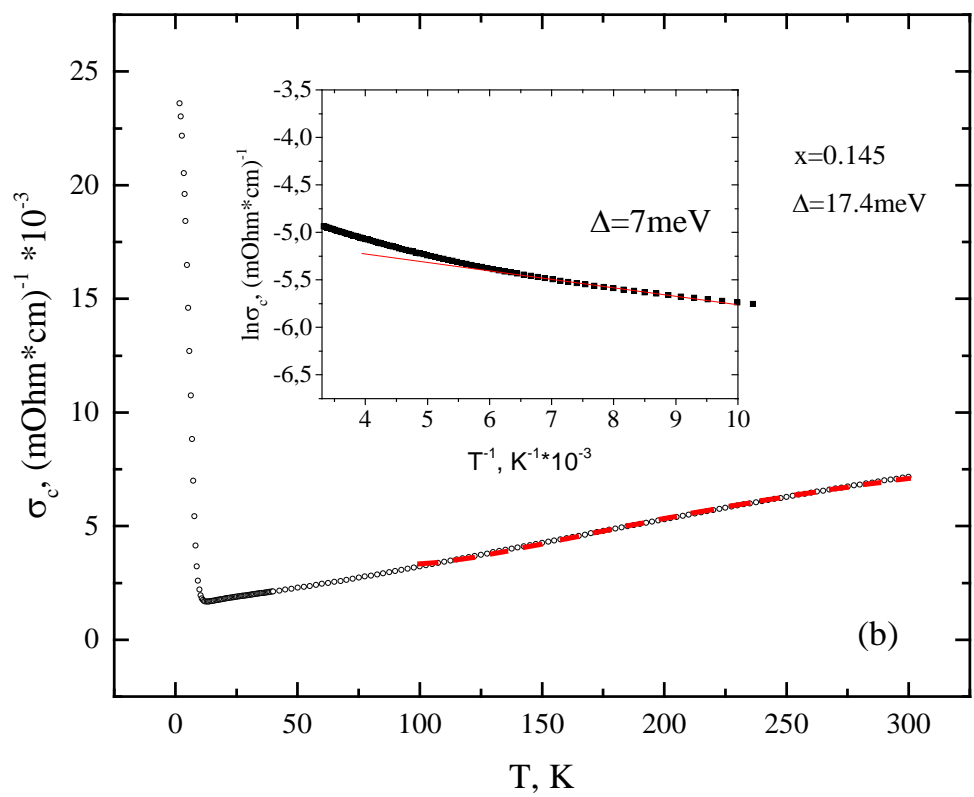

Fig2b

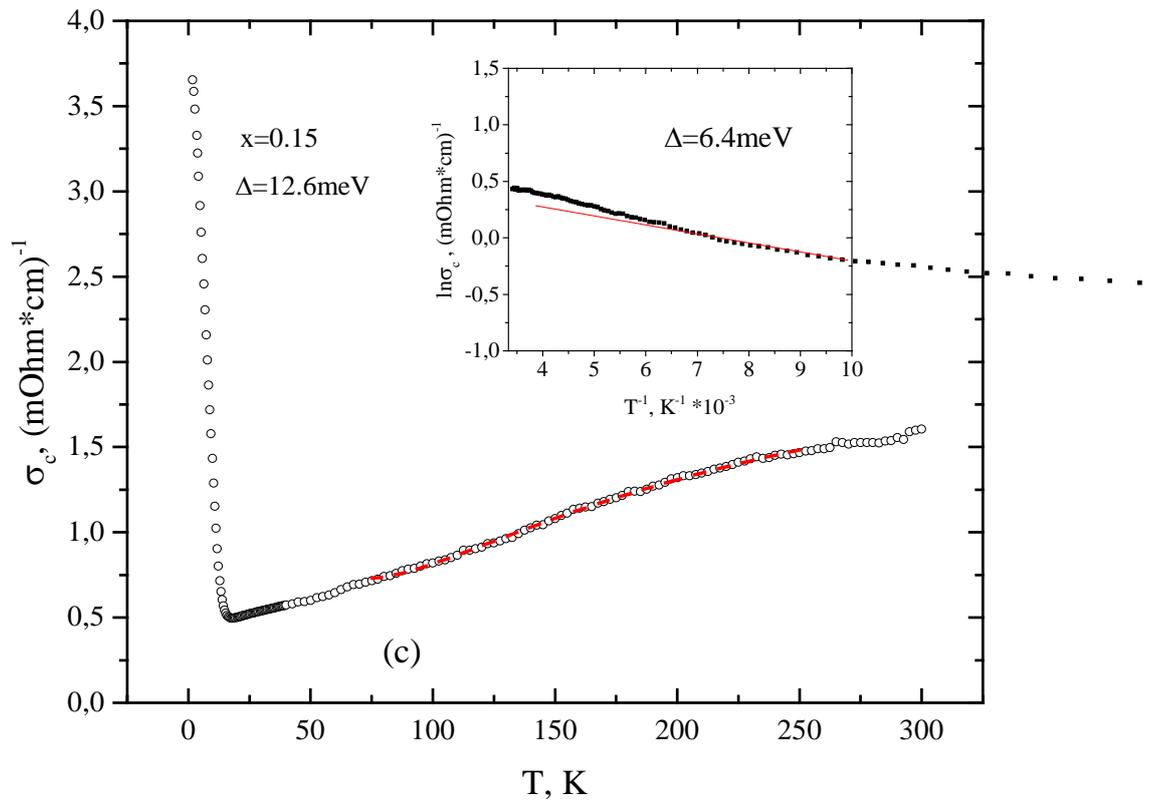

Fig2c

Title of figures:

Fig. 1. Temperature dependencies of in-plane, $\rho_{ab}$, and out-of-plane, $\rho_c$, resistivities of $Nd_{2-x}Ce_xCuO_4/SrTiO_3$ films at different doping and optimal annealing for $x = 0.135(a)$; $0.145(b)$ and $0.15(c)$. The solid lines are the best fits of Eqs (2) and (3) to our experimental data on $\rho_{ab}(T)$. The dashed lines are fitting of the data on $\rho_c(T)$ by function $a + b/T$. Insets: coefficient of resistivity anisotropy, $\rho_c/\rho_{ab}$, as a function of $T$ for each of the films.

Fig. 2. Temperature dependence of $c$-axis conductivity for $x = 0.135$ (a), 0.145 (b) and 0.15 (c). The dashed lines are the best fits of Eq. (10) to the data. Insets: the fitting of Eq. (8) to experimental $ln\sigma_c(T)$ on $1/T$ dependencies.

Tables:

| Sample, x | $\rho_{ab}(0)$, mOhm·cm | K, mOhm·cm | $T_{ee} \times 10^{-3}$, K | $\varepsilon_A$, meV | $\Delta$, meV |
|---|---|---|---|---|---|
| 0.135 | 0.47 | 37.85 | 3.17 | 1.7 | 28.6 (8.1*) |
| 0.145 | 0.34 | 56.06 | 4.43 | 2.4 | 17.4 (7.2*) |
| 0.15 | 0.03 | 328.16 | 36.8 | 2.2 | 12.6 (6.4*) |

Title of tables:

Table 1. The values of parameters obtained from a fitting of Eqs (2), (3), (4) and (10) to the corresponding experimental data.
    *The values from a fitting of $\sigma_c(T)$ by Eq. (8).